\documentclass{iopconfser}
\usepackage{amsmath,amssymb}
\usepackage{orcidlink}
\usepackage{hyperref}
\usepackage{bm}
\newcommand{\sgra}{Sgr\,A$^\star$}

\begin{document}

\title{Fundamental Physics with Pulsars around Sagittarius A$^\star$}
\author{Lijing Shao\,\orcidlink{0000-0002-1334-8853}$^{1,2}$ 
    and Zexin Hu\,\orcidlink{0000-0002-3081-0659}$^{1,3}$}

\affil{$^1$Kavli Institute for Astronomy and Astrophysics, Peking University,
Beijing China}

\affil{$^2$National Astronomical Observatories, Chinese Academy of Sciences,
Beijing China}

\affil{$^3$Department of Astronomy, School of Physics, Peking University,
Beijing China}

\email{lshao@pku.edu.cn}
\begin{abstract}
Searching for radio pulsars orbiting around the Galactic centre black hole (BH),
Sagittarius\,A$^\star$ (\sgra), represents a {\it holy grail} goal for
large-area radio telescopes, in particular for the Square Kilometre Array.
Follow-up timing observation of such a PSR-\sgra{} {\it binary} system with an
orbital period $\lesssim {\cal O}(1\,{\rm year})$ will bring forward a handful
of new tests on different aspects of fundamental physics that are barely
accessible with other means.  However, mass perturbation in the Galactic centre
harms the {\it gravitational cleanness} of PSR-\sgra{} systems. In order to
flexibly account for perturbations, a numerical pulsar timing model is gradually
being built, which can be used to probe the spacetime around \sgra{} BH and
study the nature of dark matter.
\end{abstract}

\section{Introduction} \label{sec:intro}

Over the past decades, astronomers have verified the existence of a supermassive
black hole (BH), Sagittarius\,A$^\star$ (\sgra), at the centre of our Galaxy, by
monitoring stellar orbits~\cite{Ghez:2008ms, Gillessen:2008qv}. The picture is
also consistent with observations of other galaxies~\cite{Kormendy:2013dxa}.
\sgra{} is a true treasure for various aspects of astrophysics and fundamental
physics~\cite{2023arXiv231011912S}, with its shadow image being an outstanding
example~\cite{EventHorizonTelescope:2022wkp}. In order to fully explore the
potential of \sgra{}~\cite{Kramer:2004hd, Bower:2018mta, 2019BAAS...51c.438B}, 
astronomers put great efforts to search for radio pulsars closely orbiting it.
Using information from a wide range of multi-wavelength
measurements, $\sim10^3$ pulsars were predicted to exist in the central
parsec~\cite{Wharton:2011dv}.  Although none of past surveys detected radio
pulsars yet with a desired orbital period $P_{\rm b} \lesssim {\cal O}(1\,{\rm
year})$ around \sgra{}, this fact is by now reasonably accounted for with
uncertainties from population estimates and survey specifics (see e.g.,
Refs.~\cite{Liu:2021ziv, EventHorizonTelescope:2023atv, Desvignes:2025hkk}).
There is still vast room for crucial discoveries of radio pulsars, in
particular with the Square Kilometre Array (SKA) to be operational soon in the
southern hemisphere~\cite{Kramer:2004hd, Shao:2014wja, Weltman:2018zrl}.

In order to search, as well as perform subsequent timing observation, for a
radio pulsar closely orbiting around \sgra{}, a precise timing model is needed.
Because of the relativistic nature of the pulsar orbit in the BH spacetime, the
first post-Newtonian (PN) order approximation used in the usual binary pulsar
timing models (see e.g., Refs.~\cite{Freire:2024adf, Hu:2023vsq}) is not enough.
Given the timing precision, some higher-order PN terms play a significant role
in orbital dynamics and ligh propagation.  In particular, the rotation of
\sgra{} introduces notable frame dragging of the orbit, resulting in
characteristic timing signals that can be used to extract the \sgra{}
spin~\cite{Wex:1998wt}. However, mass distribution in the Galactic centre
region, influencing the trajectory of the pulsar as well as radio light
propagation, hinders a {\it gravitationally clean} inference of the spacetime of
\sgra{}~\cite{Liu:2011ae, Psaltis:2015uza}. 

Psaltis {\it et al.}~\cite{Psaltis:2015uza} have conducted simulations with
timing data collected only around the periastron passages where effects from
mass perturbations are small, and compared pulsars' measurement ability with
that of stars and BH shadow imaging.  However, they also showed that if the full
orbit can be used, the precision will be much better. In order to properly
account for mass perturbations, we are gradually building a numerical 
pulsar-timing model specifically for tight PSR-\sgra{} binaries whose $P_{\rm b}
\lesssim {\cal O}(1\,{\rm yr})$. Here we review the model in
Sec.~\ref{sec:model} and highlight a few important science cases in
Sec.~\ref{sec:app}.

\section{A Numerical Timing Model} \label{sec:model}

In Einstein's general theory of relativity, the only possible vacuum solution
for an uncharged, stationary, rotating BH is the Kerr metric. It is called the
no-hair theorem. Such a Kerr BH is only described by two quantities, its mass
$M$ and spin $\bm{S}$. Other higher-order mass moments are totally determined by
them. For example, the quadrupole moment is given by $Q = - S^2 / Mc^2$, where
$S \equiv |\bm{S}|$ and $c$ is the speed of light. In terms of dimensionless
quantities, $\chi \equiv cS/GM^2$ and $q\equiv c^4 Q/G^2M^3$, the relationship
reads $q = -\chi^2$. Therefore, if $M$, $\bm{S}$, and $Q$ can be {\it
independently} measured, one can perform a test of the no-hair
theorem~\cite{Kramer:2004hd}.  Besides the spin-induced frame-dragging
effects~\cite{Wex:1998wt}, quadrupole moment of \sgra{} introduces timing
residuals that are usually larger than the timing precision of radio
pulsars~\cite{Liu:2011ae, Psaltis:2015uza, Bower:2018mta}.  Therefore, the
quadrupole moment can also be inferred. Overall, pulsar timing, benefitting from
the stable rotation of pulsars and the precise measurement of the times of
arrival at the site of large-area radio telescopes, provides a neat way to {\it
independently} measure $M$, $\bm{S}$, and $Q$~\cite{Wex:1998wt, Liu:2011ae,
Psaltis:2015uza}.

\begin{figure*}
	\centering
	\includegraphics[width=13cm]{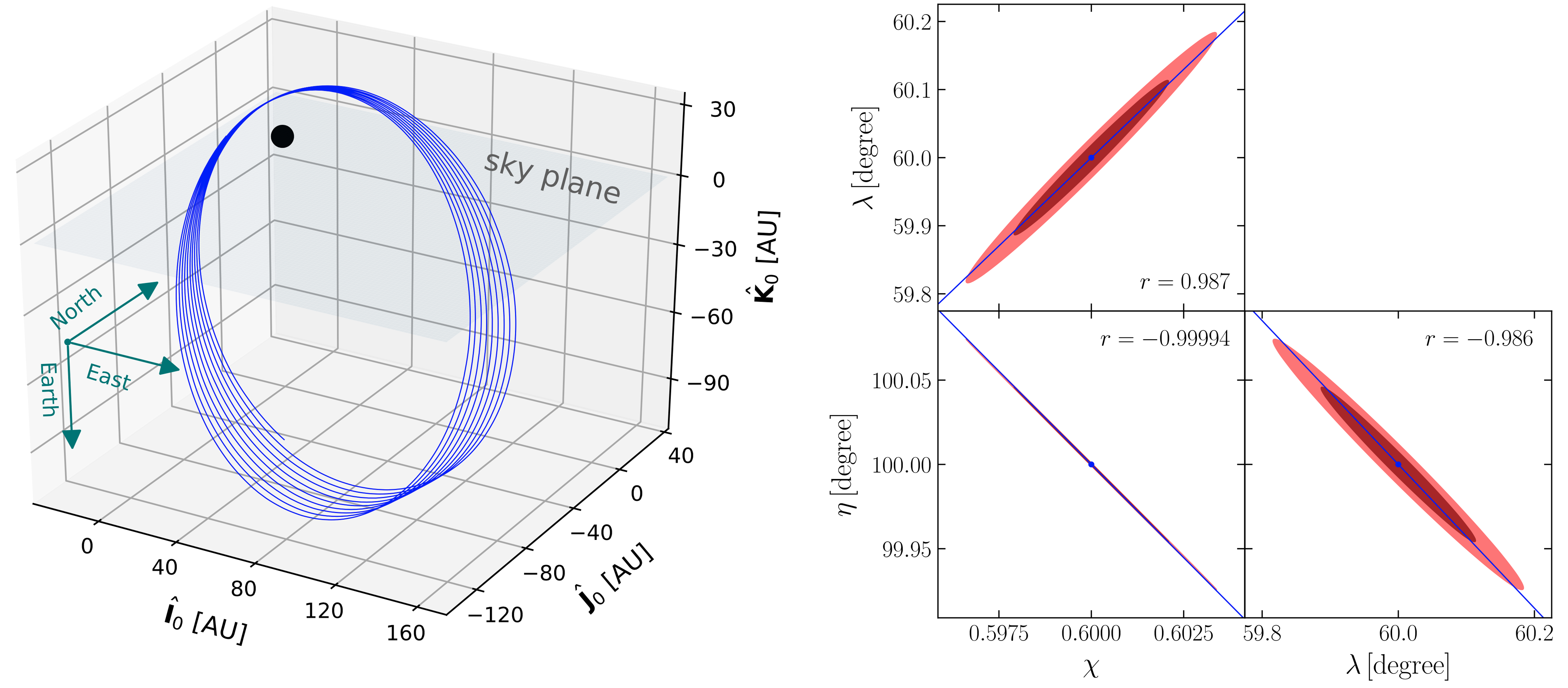}
	\caption{ Example of pulsar trajectory around \sgra{} with $P_{\rm
	b}=0.5\,{\rm yr}$, $e=0.8$, and $\chi=0.6$ ({\it Left}); Correlations
	between three spin parameters ({\it Right}). Reproduced from
	Ref.~\cite{Hu:2023ubk} with
	\href{https://creativecommons.org/licenses/by/4.0/}{License~CC~BY~4.0}.
	\label{fig}}
\end{figure*}

As mentioned in Sec.~\ref{sec:intro}, in order to make full use of the
pulsar-timing data even in presence of mass perturbations, we resort to
numerical integration of the pulsar orbit and light
propagation~\cite{Hu:2023ubk}. Our orbital trajectory of the pulsar (see the
left panel of Fig.~\ref{fig}) is based on solving the PN equation of motion,
\begin{equation}
	\ddot{\boldsymbol{r}} \equiv \frac{\mathrm{d}^2 \boldsymbol{r}}{\mathrm{~d}
	t^2}=\ddot{\boldsymbol{r}}_{\mathrm{N}}+\ddot{\boldsymbol{r}}_{1
	\mathrm{PN}}+\ddot{\boldsymbol{r}}_{\mathrm{SO}} +
	\ddot{\boldsymbol{r}}_{\mathrm{Q}}+\ddot{\boldsymbol{r}}_{2
	\mathrm{PN}}+\ddot{\boldsymbol{r}}_{2.5 \mathrm{PN}}+\cdots \,, \label{eq}
\end{equation}
where $\bm{r}$ is the relative coordinate vector of the orbit in the harmonic
gauge, and $\ddot{\bm{r}}_{\rm N} = -GM \bm{r} / r^3$ is its Newtonian
acceleration, while $\ddot{\bm{r}}_{\rm 1PN}$, $\ddot{\bm{r}}_{\rm 2PN}$, and
$\ddot{\bm{r}}_{\rm 2.5PN}$ are the relativistic accelerations of two point
masses at different PN orders; $\ddot{\bm{r}}_{\rm SO}$ and $\ddot{\bm{r}}_{\rm
Q}$ are the accelerations caused by the BH spin and quadrupole moment,
respectively. Concrete expressions of these terms can be found in
Ref.~\cite{Hu:2023ubk}. Light propagation is also taken care of with proper
approximation. Following steps of~Damour and Deruelle~\cite{Damour:1986ads}, a
PSR-\sgra{} timing model is built by coherently combining orbital trajectory
and light propagation~\cite{Hu:2023ubk}.

Given timing precision and observational cadence, we use the Fisher matrix to
forecast the parameter uncertainties~\cite{Hu:2023ubk}.  There was  a
semi-analytic timing model for PSR-\sgra{} systems in the pulsar community which
resembles the Damour-Deruelle timing model~\cite{Liu:2011ae, Psaltis:2015uza}.
We simulated PSR-\sgra{} systems with different orbital periods ($P_{\rm b}$)
and orbital eccentricities ($e$), and got consistent results with
Refs.~~\cite{Liu:2011ae, Psaltis:2015uza} on the uncertainties of $M$, $\bm{S}$,
and $Q$, thus verifying  our numerical timing model.

From the parameter estimation, we can see from the right panel of Fig.~\ref{fig}
that the three spin parameters---the (dimensionless) magnitude $\chi$ and its
two positional angles $\lambda$ and $\eta$---are highly correlated, therefore
relatively poorly measured.  We noticed that, from population estimates, the
probability of discovering two moderate pulsars with $P_{\rm b} \sim 2 \mbox{--}
5\,{\rm yr}$ is larger than the probability of discovering one pulsar in a tight
orbit with $P_{\rm b} \lesssim 0.5\,{\rm yr}$~\cite{Hu:2024blq}. Therefore, it
is worth studying the combination of two moderate pulsars. Because the
orientation of two orbits is likely to incline with respect to each other, it
largely breaks the degeneracy of three spin parameters, and achieves a
comparable measurement of the BH spin as one may get from a tight orbit~\cite{Hu:2024blq}.

\section{Applications to Fundamental Physics} \label{sec:app}

With such a numerical model at hand, it opens the possibililty to study an array
of tests on different aspects of fundamental physics.  Below we give some recent
examples.


{\bf Small-scale Dark Matter (DM) Distribution.} The timing model is flexible to
include mass perturbations in Eq.~(\ref{eq}). As a start, a static spherical
distribution was investigated~\cite{Hu:2023ubk}.  Such a mass distribution
around the \sgra{} BH was predicted by some DM models, and is called the {\it DM
spike}. In this scenario, a pulsar travels inside this DM spike, and its
trajectory can be solved numerically with the addition of an extra acceleration
term to Eq.~(\ref{eq}). The existence of DM spike also deflects the propagation
of radio signals and contributes to the usual Shapiro delay.  After taking a
spike profile, say, a (modified) power-law distribution with its magnitude and power index
as free parameters, we  applied the Fisher matrix to our model and measured the
profile parameters. Once such a PSR-\sgra{} system is discovered, it will be the
first experiment to detect DM distribution on a scale as small as
milli-parsec~\cite{Hu:2023ubk}.


{\bf Yukawa Gravity.} When the gravity is mediated with a massive degree of
freedom, a common way to parameterize the leading-order gravitational potential
is $\varphi(r) = -GM \big[ 1 + \alpha \exp(-r/\Lambda)  \big] / \big[ (1+\alpha)
r \big]$, where $\alpha$ is the strength and $\Lambda$ gives the lengthscale of
suppression, which is inversely related to the mediator mass $m_g$.  From
$\varphi(r)$ one obtains a modified {\it Newtonian} acceleration term,
$\ddot{\bm{r}}_{\rm N}^{\rm (modified)}$, for
Eq.~(\ref{eq})~\cite{Dong:2022zvh}. With this modification, a modified pulsar
trajectory can be obtained, and combining with the Fisher matrix, we obtained
the measurement precision of $\alpha$ for different values of $m_g$. Simulations
showed that, in the relevant $m_g$ range, radio pulsars will outperform the
monitoring of stellar objects~\cite{Ghez:2008ms, Gillessen:2008qv} in this
Yukawa gravity test by several orders of magnitude~\cite{Dong:2022zvh}.


{\bf Vector-tensor Gravity.} We can also combine our numerical timing model with
specific gravity theories.  In a class of Lorentz-violating vector-tensor theory
where an extra vector is nonminimally coupled to gravity, the so-called
bumblebee gravity theory~\cite{Bailey:2006fd}, a complete set of solution for spherical
symmetric, static spacetime was obtained by Xu {\it et al.}~\cite{Xu:2022frb}.
We used the metric solution to obtain pulsar trajectory around \sgra{}, and
investigated the constraints that can be put from realistic
timing~\cite{Hu:2023vsg}. Our results showed that the limits on the vector
charge will be a factor of ${\cal O}(10^2)$ times better than that was obtained
using the BH imaging~\cite{Xu:2023xqh}.


{\bf Long-range Fifth-fore from DM.} Other than the small-scale properties of DM
distribution mentioned earlier, pulsars around \sgra{} can also be used to put
limits on fundamental interactions of DM with ordinary matter.  While gravity is
geometry in the general relativity,  the interaction of DM and ordinary matter
might have other forms besides the gravitational interaction. Short-range
fifth-force, say, with nucleons, is commonly considered, but long-range forces
are seldomly investigated.  If there is a long-range fifth-force between DM and
ordinary matter, two components of a binary system in the DM halo fall
differently, {\it apparently} violating the equivalence principle if this
long-range fifth-force is ignored~\cite{Shao:2018klg}. Borrowing the idea of
using evolution of the eccentricity vector to test the equivalence principle
with binary pulsars from Damour and Sch\"afer~\cite{Damour:1991rq}, the strength
of fifth-force was constrained to be smaller than $\sim 1\%$ of the
gravitational interaction, otherwise it has already been conflicting with the timing
measurement of PSR~J1713+0747~\cite{Shao:2018klg}.  The high-density DM spike
mentioned before could further enhance the test. When proper binary pulsars are
found within $\sim 10\,{\rm pc}$ to the Galactic centre, the test of DM
long-range fifth-force will improve by several orders of magnitude~\cite{Shao:2018klg}.

\section{Summary}

In recent years we have witnessed a flourishing of strong-field gravity tests
from pulsar timing, gravitational waves, and BH shadow
imaging~\cite{EventHorizonTelescope:2022xqj}. Einstein's general relativity has
passed all the tests with flying colors. However, as a fundamental theory in
physics, it is never a safe period to stop being examined with empirical data.
Discovering radio pulsars in the vicinity of \sgra{}, likely with the SKA, might
provide new opportunities to perform fundamental tests of gravity, as well as
tests of other fundamental aspects of physics, say, the nature of dark matter. In order to
fully embrace the era, we are gradually building a flexible, numerical pulsar
timing model for PSR-\sgra{} systems, that is capable of testing generic and
specific gravity theories, dark matter properties, and more.

\section*{Acknowledgements}

This work was supported by the National SKA Program of China (2020SKA0120300),
the National Natural Science Foundation of China (124B2056), the Beijing Natural
Science Foundation (1242018), the Max Planck Partner Group Program funded by the
Max Planck Society, and the High-Performance Computing Platform of Peking
University.
 
\bibliographystyle{iopart-num}
\bibliography{refs}

\end{document}